\documentclass[aps,prl, twocolumn, nofootinbib, longbibliography]{revtex4-2}

\makeatletter
\g@addto@macro\bfseries{\boldmath}
\makeatother

\usepackage{amsmath,amssymb,graphicx,dcolumn,bm,hyperref,float}
\usepackage{microtype}
\hypersetup{ colorlinks=true, linkcolor=red, filecolor=red,  urlcolor=red, citecolor = red}
\usepackage{xcolor}
\usepackage{csquotes}
\usepackage{tikz}
\usepackage{amsthm}
\usepackage[caption=false]{subfig}
\usepackage{braket}
\usepackage{yfonts}

\graphicspath{{./figures/}}

\begin{document}

\title{Critical behavior of dirty parafermionic chains}

\author{Akshat Pandey}
\email{akshatp@stanford.edu}
\affiliation{Department of Physics, Stanford University, Stanford, CA 94305, USA}

\author{Aditya Cowsik}
\affiliation{Department of Physics, Stanford University, Stanford, CA 94305, USA}

\date{\today}

\begin{abstract}
    A family of $\mathbb Z_n$-symmetric non-Hermitian models of Baxter was shown by Fendley to be exactly solvable via a parafermionic generalization of the Clifford algebra. 
    We study these models with spatially random couplings, and obtain several exact results on thermodynamic singularities as the distributions of couplings are varied.
    We find that these singularities, independent of $n$, are identical to those in the random transverse-field Ising chain; correspondingly the models host infinite-randomness critical points. 
    Similarities in structure to exact methods for random Ising models, a strong-disorder renormalization group, and generalizations to other models with free spectra, are discussed.
\end{abstract}

\maketitle

\textit{Introduction.---} Consider an open chain of length $L$ with an $n$-dimensional Hilbert space on each site. With each site associate analogues of Pauli $Z$ and $X$ matrices, defined as
\begin{equation}
    Z = \begin{bmatrix}
        1 & & & \\
         & \omega & & \\
         & & \dots &\\
         & &  & \omega^{n-1} 
    \end{bmatrix}, \quad X = 
    \begin{bmatrix}
        0 & & & 1 \\
        1 & 0 & & \\
         & 1 & \dots & \\
         & & & 0
    \end{bmatrix}
\end{equation}
respectively. Here $\omega = e^{2 \pi i/n}$. The non-Hermitian Hamiltonian
\begin{equation}
    H = - \sum_{j=1}^{L-1}  J_j Z_j^\dagger Z_{j+1} - \sum_{j=1}^{L} h_j X_j
\end{equation}
is then a natural $\mathbb{Z}_n$-symmetric generalization of the transverse-field Ising chain (TFIC). It was introduced in relation with the integrable chiral Potts model by Baxter~\cite{Baxter1989PhysLett, Baxter1989JStatPhys}, who demonstrated the remarkable fact that $H$ has $n^L$ eigenvalues which can be written as
\begin{equation}\label{eq:spectrum}
    E_{\{n_k\}}  = -\sum_{k=1}^L \omega^{n_k} \varepsilon_k,
\end{equation}
where each $n_k \in \{0\dots n-1\}$ is chosen independently. The single-particle energies $\varepsilon_k$ are related to the roots of a particular polynomial whose construction we will discuss. An understanding of the ``free parafermionic'' nature of the spectrum was achieved by Fendley~\cite{FendleyFreeParafermions}, who provided an explicit construction of the parafermionic analogue of raising and lowering operators associated with each single-particle level. A Jordan-Wigner-like transformation~\cite{FradkinKadanoff}, unlike in the TFIC, does not make this task at all trivial: the aforementioned parafermionic operators are highly nonlinear in these Jordan-Wigner parafermions. 

While a number of observables in the homogeneous ($J_j = J, h_j = h$ for all $j$) version of the model, especially in the vicinity of $h = J$, have been calculated~\cite{Alcaraz2017Critical, Alcaraz2018Boundary, Batchelor2019FeynmanHellman, Alcaraz2020MultispinShort, Alcaraz2020MultispinLong, Alcaraz2021MassGapNumerics, BatchelorExceptionalPoints}, that the model is in principle exactly solvable for inhomogeneous couplings has not been exploited. This is the goal of the present paper.

We have two entwined motivations. First, given the scarcity of theoretical tools available to analyse strongly disordered quantum critical points in a controlled manner---the strong-disorder RG being the most versatile but, by its nature, limited to cases where infinite-randomness scaling is achieved and errors accumulated in the approach to said scaling are unimportant---we wish to understand the critical properties of any disordered models on which exact methods can be brought to bear. As far as we know, the only problem on which such progress was made is the disordered TFIC, or equivalently the stripy-random classical 2d Ising model~\cite{McCoyWuThermodynamics, McCoyWuCorrelations, McCoyBoundary, McCoyGeneralDistribution, ShankarMurthy}. Second, while the free parafermionic models have some features in common with the TFIC, the method of solution is more intricate, so it is not \textit{a priori} obvious that the free parafermionic character of the problem is useful in practice in the presence of disorder. We show, however, that much of the structure of the Ising papers referred to above carries over in a non-trivial way to the parafermionic models. These rather attractive relations may be useful for an understanding of the algebraic structures underlying ``free'' statistical mechanics in general, and indeed for the construction of new disordered critical models.

We leverage these results to analyze the singularity structure of the ``ground-state energy'' of $H$, and find that disorder washes out the difference between the critical behaviors among different $n$.

\textit{The ground state energy.---} The recipe for finding the single-particle energies $\varepsilon_k$ of \eqref{eq:spectrum} is as follows~\cite{Baxter1989PhysLett, FendleyFreeParafermions}. Write the couplings of the Hamiltonian as $(b_1, b_2, b_3, \dots, b_{M}) \equiv (h_1, J_1, h_2, \dots, h_L)$, so that $M = 2L-1$. Define a sequence of polynomials $P_m(z)$ using the relations
\begin{gather}
    P_0(z) = P_{-1}(z) = 1; \nonumber\\
     P_m(z) = P_{m-1}(z) - z b_m^n P_{m-2}(z),\quad  m \in \{1 \dots M\}.\label{eq:recursion}
\end{gather}
If all $b_m > 0$ (we assume this for the rest of the paper), $P_M(z)$ has $L$ positive real zeros $z_k, k \in \{1\dots L \} $. Then $\varepsilon_k = z_k^{-1/n}$. We are interested in the state in which all levels are filled in the $n_k = 0$ manner, $E_0 = -\sum_k \varepsilon_k$.
This is of course the analogue of the ground state energy. Its importance stems from the fact that it is at the edge of the many-body spectrum in the complex plane (up to multiplication by $\omega$), and it has singularities as the distributions of couplings are varied; it is these singularities that we analyse.

\begin{figure}
    \centering
    \includegraphics[width=0.7\columnwidth]{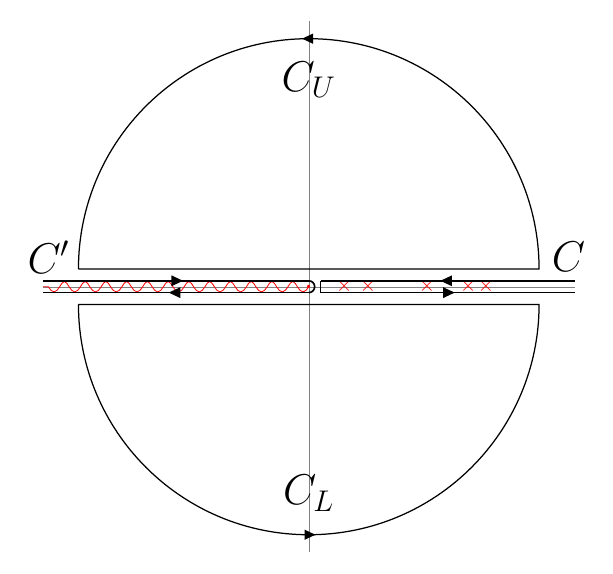}
    \caption{Singularities in the complex $z$ plane of the integrand in \eqref{eq:contourC}---crosses are poles and the wavy line is a branch cut---and contours involved in its evaluation.}
    \label{fig:contours}
\end{figure}

We now show a simple formula for $E_0$ in terms of $P_M$ which we use for all subsequent developments. We start by writing
\begin{equation}\label{eq:contourC}
-E_0 = \frac{1}{2\pi i}\int_C dz\,  z^{-1/n} \frac{P_M'(z)}{P_M(z)},
\end{equation}
where $C$ is the contour shown in Fig.~\ref{fig:contours}. This converts the zeros $z=z_k$ of $P_M$ to poles whose residues contain the requisite factors of $\varepsilon_k$. We can append the contours $C_U, C_L$ since they enclose no singularities. Their semicircular parts vanish (in the limit of large radii), so that we can integrate along $C'$ which straddles the branch cut. Integrating by parts, and changing variables to $\Omega = |z|^{-1/n}$, we obtain
\begin{equation}~\label{eq:E0formula}
    -E_0 = \frac{1}{\pi} \sin\frac{\pi}{n} \int_0^\infty d\Omega \, \ln P_M(-\Omega^{-n}).
\end{equation}
This formula reflects the free nature of the model: the partition function factorizes into frequencies which do not talk to each other. For $n=2$, $E_0$ is an appropriate limit of the free energy of a 2d Ising model which is inhomogeneous in one direction and homogeneous in the other: see the works by McCoy and Wu~\cite{McCoyWuThermodynamics}, and Shankar and Murthy~\cite{ShankarMurthy}. Indeed the strategy in these works was to factorize the partition function into wavevectors in the clean direction (which for us is time). The procedure, however, relied on manipulations particular to the Ising model (using Pfaffians~\cite{McCoyWuBook}, and Jordan-Wigner fermions~\cite{SchultzMattisLieb}, respectively). It is interesting that the weaker notion of freeness in the models here (and others which we will comment on) is sufficient to produce the same structures. The connections are also of practical value, as we will demonstrate presently.

\textit{Singularities of $E_0$.---} We now study the non-analyticities $E_0$ exhibits as a function of the distributions of couplings. Henceforth all $h_j$ will be set equal to $1$, while $J_j$ will be independent random variables drawn from a distribution $P(J)$. The precise forms of the distributions do not affect the physics we discuss (as long as tails in the distributions are not too heavy~\cite{fisher1995critical}).

Following the Ising work of Shankar and Murthy~\cite{ShankarMurthy}, we interpret $ P_M(-\Omega^{-n})$ as the partition function of a 1d \textit{classical} Ising model with random fields. The recursion \eqref{eq:recursion} can be written as 
\begin{equation}\label{eq:matrixrecursion}
    \begin{bmatrix}
        P_m(-\Omega^{-n})\\ P_{m-1} (-\Omega^{-n})
    \end{bmatrix}
    = 
    \begin{bmatrix}
        1 &  b_m^n \Omega^{-n}\\
        1 & 0
    \end{bmatrix}
    \begin{bmatrix}
        P_{m-1}(-\Omega^{-n}) \\ P_{m-2}(-\Omega^{-n})
    \end{bmatrix}.
\end{equation}
Multiplying the matrices associated with $h_{j+1} (= 1)$ and $J_{j}$ yields
\begin{equation}\label{eq:Tdefinition}
    T_j (\Omega) = \begin{bmatrix}
        1 +  \Omega^{-n} & J_j^n \Omega^{-n} \\
        1 & J_j^n \Omega^{-n}
    \end{bmatrix}.
\end{equation}
Consider now a classical Ising model defined by the Hamiltonian $-\beta H_I = \sum_j [ \mathcal{E}_j + \mathcal{J} (\sigma_j \sigma_{j+1}-1) + \mathcal{H}_{j} \sigma_j + \mathcal{H}'_{j}\sigma_{j+1}]$. The transfer matrix that assembles these Boltzmann weights is clearly $T_j$ if we identify $\mathcal{E}_j = (1/2) \ln[J_j^n \Omega^{-n}(1+\Omega^{-n})]$, $\mathcal{J} = \mathcal{H}_j = (1/4) \ln(1 +  \Omega^{-n})$ and $\mathcal{H}'_j = (1/4) \ln(1 + \Omega^{-n}) - (1/2) \ln(J_j^n \Omega^{-n})$. The fields can be redistributed among sites so that all $\mathcal H'_j $ vanish, and $\mathcal{H}_j = -(1/2) \ln(1 +  \Omega^{-n}) + (1/2) \ln(J_j^n \Omega^{-n})$---we have also multiplied the field by a minus sign for later convenience. This puts the Hamiltonian into a more standard random-field Ising form. This sort of ``gauge'' freedom in the transfer matrix (namely the ability to modify couplings as long as $\mathcal{J}$, and all $\mathcal{H}_j + \mathcal{H}'_{j-1}$, up to a $j$-independent sign, remain unchanged---equivalently to insert factors of $1 = AA^{-1}$ for appropriate matrices $A$) is useful to make a number of connections between quantities most naturally expressed as products of different families of random matrices that ultimately yield the same polynomials. Still different factorizations will be exploited in the next two sections.

Returning to the problem at hand, to find the ground state energy density $E_0/L$, we have to integrate a quantity of the form $f(\Omega) = (1/L) \ln \bra{v} \prod_j T_j(\Omega) \ket{w}$, where the boundary vectors $v$ and $w$ can be read off $\eqref{eq:recursion}$, and the product goes down in $j$ rightward. Each $f(\Omega)$ is a 1d random-field Ising model's free energy density, as we have just explained. These are self-averaging quantities in the $L\to \infty$ limit, and so is $E_0/L$. Singularities in $E_0/L$ are inherited from the approach to singularities in $f(\Omega)$ in the zero-temperature limit, which translates to $\Omega \to 0$. Shankar and Murthy used results from Refs.~\cite{DerridaHilhorst1983, NL} to extract these singularities, and our calculation now closely follows theirs. Let us suppose $P(J)$ has support for $J \in [J_1, J_2]$. We expect Griffiths-type singularities when $J_1$ crosses $1$ and when $J_2$ crosses $1$. 
These correspond to points when the $ \Omega \to 0$ classical chain goes from having all up-pointing fields to having arbitrarily long regions where the field points down, and vice versa.  Now let us specialize to $P(J) = [\delta(J-J_1)+\delta(J-J_2)]/2$, $J_1 \ll 1$ and $J_2 = 1+t_G/n$ with $0 <t_G\ll 1$, to focus on the vicinity of the latter of these Griffiths singularities, viz. the paramagnetic one. In this regime (from \cite{DerridaHilhorst1983, ShankarMurthy})  $ f(\Omega)$ has a singular part $e^{-4\alpha \mathcal{J}},$ with the exponent $\alpha$ defined by $ \overline{e^{2\alpha \mathcal{H}}} = 1$, for $\Omega$ small enough that not all $\mathcal{H}$ are the same sign. (Overlines denote disorder averages.) The latter relation gives $\alpha \approx \ln 2/(t_G - \Omega^n)$, and so we are to integrate $e^{-4\alpha \mathcal{J}}$ up to $\Omega < t_G^{1/n}$. We get that the singular part of $E_0/L$ is proportional to
\begin{equation}
    \int_0^{t_G^{1/n}} d\Omega \, \,  \Omega^{n\ln 2/(t_G-\Omega^n)} \sim t_G^{c/t_G} ,
\end{equation}
where $c$ is a constant. There will be an identical singularity as $J_1$ approaches $1$ from below.

The other interesting point in the classical chain is the one where the average field switches direction, $\overline{\mathcal H} = 0$. This corresponds to the ferromagnet-paramagnet transition in the quantum chain, which occurs when $\overline{\ln h} =\overline{\ln J}$. Let us now consider the behaviour in the vicinity of this point. Shankar and Murthy spliced together results from Refs.~\cite{DerridaHilhorst1983, NL} to argue that all the singularities of $f$ in the vicinity of $\overline{\mathcal{H}} = 0$ are captured by the function  $f\sim -\overline{\mathcal H} \coth(2\alpha \mathcal{J})$ where $\alpha$ is defined as above. Again assuming all $h = 1$ and defining $t=- (n/2) \overline{\ln J}$, $\alpha \sim -\overline{\mathcal H} \sim t + \Omega^n $, where we have dropped, and will continue to drop, a number of constants. The energy then has the same singularity as
\begin{equation}
    \int_0^{\epsilon} d\Omega \,  (t + \Omega^n) \coth[(t + \Omega^n) \ln\Omega^{-n}].
\end{equation}
with any $\epsilon < 1$ (the non-analyticity comes from the integral near zero). Shankar and Murthy's observations on an integral of this form with $n=2$ in fact generalize to all $n$~\cite{ShankarMurthy}: to obtain the most important piece of the singularity we can drop the $\Omega^n$ additive pieces, so that the integral becomes $\int_0^\epsilon d\Omega \, t \coth(nt \ln \Omega^{-1})$. Up to a trivial scaling by $n$, this is the same function of $t$ for all $n$. It is also infinitely differentiable yet non-analytic at $t=0$~\cite{McCoyWuThermodynamics, ShankarMurthy}.
% , and yield that the most singular part goes like
% \begin{equation}
%     \sum_{m=0}^\infty  \frac{1}{(2m)!} \Gamma(2m) (2nt)^{2m} B_{2m}  , 
% \end{equation}
% where $B_{2m}$ denotes the Bernoulli numbers. Therefore the energy as a function of $t$ is infinitely differentiable at $t=0$, but the clear divergence of the series implies that it is not analytic at this point.

The energy $E_0$ therefore contains three infinitely differentiable singularities as the distributions of $h$ and $J$ are tuned across each other, of the same form as those in the Ising ($n=2$) case~\cite{McCoyWuThermodynamics, ShankarMurthy}. This should be contrasted with the clean Hamiltonians, in which one finds a usual critical exponent $\alpha = 1-2/n$~\cite{Alcaraz2017Critical}.

\textit{Boundary magnetization.---}  While introducing magnetic field ($Z$  or $Z^\dagger$) type terms in the bulk destroys the free-parafermion solvability of the model, a field on the boundary does not. To be explicit, the Hamiltonian
\begin{equation}
    H_{\mathfrak H} = H - \mathfrak H Z_L^\dagger \quad (\mathfrak H > 0)
\end{equation}
is just as solvable as $H$. This is because $Z_L^\dagger$ obeys the same algebra with $X_L$ as two adjacent terms in $H$, e.g. $X_j$ and $Z^\dagger_{j-1}Z_j$. One simply introduces $b_{M+1} = \mathfrak H$ and works with the polynomial 
\begin{equation}
    P_{M+1}(z) = P_M(z) - z \mathfrak H ^n P_{M-1}(z).
\end{equation}
Note that this has the same degree as $P_M(z)$---the number of single-particle levels remains the same. 
The energy is given by \eqref{eq:E0formula} with $M$ replaced by $M+1$. We define the boundary ``magnetization'' 
% $\mathfrak M(\mathfrak H) = \langle Z_{L}^\dagger \rangle$
in the corresponding state in the presence of the boundary field  as  $\mathfrak M(\mathfrak H) =-dE_0(\mathfrak H)/d\mathfrak H$ 
which, following some trivial manipulations, can be written as
\begin{equation}
    \mathfrak M(\mathfrak H) = \frac{1}{\pi} \sin \frac{\pi}{n} \int_0^\infty d\Omega \, \frac{n \mathfrak H ^{n-1}}{\Omega^n r_M + \mathfrak H ^n },
\end{equation}
where $r_m = P_m(-\Omega^{-n})/P_{m-1}(-\Omega^{-n})$ is a random variable that depends on all the $J_j$ that $P_m$ contains. Unless the chain is clean it fluctuates as a function of $m$---see \eqref{eq:matrixrecursion}---even as $m \to \infty$, which is the limit of interest. However, in this limit it will approach a stationary distribution~\cite{ McCoyWuThermodynamics}. We will be particularly interested in the spontaneous magnetization, $\lim_{\mathfrak H \to 0^+} \mathfrak M(\mathfrak H)$.

At this point it is convenient to write our matrix recursion somewhat differently using the random-field Ising picture discussed in the previous section. It is easy to verify that if we define the matrix (cf. \eqref{eq:Tdefinition})
\begin{equation}\label{eq:tildeT}
    \tilde T_j(\Omega) = \begin{bmatrix}
        \Omega^n + 1  & J_j^n \Omega^{n/2} \\
        \Omega^{n/2} & J_j^n 
    \end{bmatrix}
\end{equation}
then 
\begin{equation}
   \left( \prod_{j = 1}^{L-1} \tilde T_j\right) \begin{bmatrix}
        \Omega^n + 1 \\ \Omega^{n/2} \end{bmatrix}= 
   \Omega^{nL}  \begin{bmatrix}
        P_M(-\Omega^{-n}) \\
        \Omega^{-n/2} P_{M-1}(-\Omega^{-n})
    \end{bmatrix}.
\end{equation}
Therefore, if we define $x = r_M \Omega^{n/2}$, and suppose that in the $M\to \infty$ limit the law of $x$ is $\nu(x)$, then the disorder-averaged boundary magnetization is given by
\begin{equation}\label{eq:boundaryMaverage}
    \overline{\mathfrak M(\mathfrak H)}  = \frac{1}{\pi} \sin \frac{\pi}{n} \int_0^\infty d\Omega\int_0^\infty dx \, \nu(x)\frac{ n \mathfrak H ^{n-1}}{\Omega^{n/2} x + \mathfrak H ^n }.
\end{equation}
This rearrangement allows a connection to the work of McCoy and Wu~\cite{McCoyWuThermodynamics}, who studied the distribution $\nu(x)$ for the form of random matrix shown in  \eqref{eq:tildeT}, with $n=2$.  
This was accomplished for a power law distribution of $J$'s, particularly  $\mu(J^n) = P(J)/(nJ^{n-1}) =  N \lambda_0^{-N} J^{n(N-1)} $ for $0 < J^n < \lambda_0$ and $0$ otherwise. The distribution $\mu(\lambda)$ is assumed to be narrow around $\lambda_0$, i.e. $N \gg 1$. The tuning parameter for the transition is therefore $\lambda_0$; the phase transition in magnetization takes place when  $t= (1/N  - \ln \lambda_{0})/2  = 0$. We work with the same distribution.
McCoy then used this $\nu(x)$ to study boundary magnetization~\cite{McCoyBoundary},  manipulating Pfaffians for a 2d Ising lattice with an extra column to get to a formula similar to \eqref{eq:boundaryMaverage}. 

McCoy and Wu's results show, \textit{mutatis mutandis}, that deviations from a clean model occur only for $|t|, \Omega^{n/2}  \lesssim N^{-2}$, motivating a change of variables $\delta \sim N^2 t, \phi \sim N^2 \Omega^{n/2},$ and correspondingly $ \mathfrak h = N^{2/n} \mathfrak H $. In the $\delta, \phi \sim 1$ regime, another change of variables $x = \lambda_0 ^{1/2}\phi e^{-q}/2$ is useful. The distribution of $q$ was determined by McCoy and Wu to be $U(q) \equiv (\phi/2)^\delta  \exp\{-\delta q - e^{q} - \phi^2 e^{-q}/4\}/(2 K_\delta(\phi))$ where $K_\delta$ is a modified Bessel function of the second kind~\cite{McCoyWuThermodynamics}. Throughout this discussion we are cavalier about factors that do not matter for the form of the singularity. We have that $N^{2/n} \overline{\mathfrak M(\mathfrak H)}  $ goes like
\begin{equation}
\int_0^\infty d\phi \int_{-\infty}^\infty dq \, \phi^{2/n-1}\left( \frac{\phi}{2}\right)^\delta \frac{e^{ -\delta q - e^{q} - \phi^2 e^{-q}/4 } \mathfrak h ^{n-1}}{ K_\delta(\phi)(\phi^{2} e^{-q} + \mathfrak  h^n)}  .
\end{equation}
In the $\mathfrak h \to 0$ limit the $\phi^2$ term in the exponent can be dropped. Making a further change of variables $\phi = \mathfrak h^{n/2}\alpha$, and plugging in the small-$\delta$ asymptotics of $K_\delta$, we have (dropping factors of order 1):
\begin{equation}
    \int_0^\infty d\alpha \int_{-\infty}^\infty dq  \frac{(\mathfrak h^{n/2} \alpha/2 )^{\delta + |\delta|}}{\Gamma(|\delta|)} \frac{\alpha^{2/n - 1}}{\alpha^2 e^{-q}+1}e^{ -\delta q - e^{q}  }
\end{equation}
so that all $\mathfrak h$ dependence is now in the factor $(\mathfrak h^{n/2} \alpha/2 )^{\delta + |\delta|}$. When $\delta > 0$ this factor vanishes as $\mathfrak h \to 0$, so there is no magnetization in the disordered phase. For $\delta < 0$, the factor is $1$, and the rest of the integral can be easily performed to yield constant factors times  $\Gamma(1/n + |\delta|)/\Gamma(|\delta|) \sim \Gamma(1/n) |\delta|$. Therefore 
\begin{equation}
     \lim_{\mathfrak H \to 0^+} \overline{\mathfrak M(\mathfrak H)}  \sim \Theta(-t) |t|^{\beta_{\text{boundary}}}; \quad \beta_{\text{boundary}} = 1.
\end{equation}
We see that the  boundary magnetization exponent is also independent of $n$.

\textit{Strong-disorder RG.---} While the focus of this paper has been on exact results, it is worth adding that our formalism allows us to make contact rather naturally with Fisher's strong-disorder RG~\cite{fisher1995critical}. Our comments in fact generalize those of Fisher's Appendix E, in which he explains the connection (for the TFIC) between his results and Shankar and Murthy's. Here one introduces yet another matrix factorization of the polynomial $P_M(\Omega^{-n})$. With matrices
\begin{equation}
    \mathcal{T}^{(h)}(h) = \begin{bmatrix}
        \Omega^{n/2}/4 - h^n \Omega^{-n/2} & -1/2 \\
        1/2 & -\Omega^{-n/2}
    \end{bmatrix},
\end{equation}
and
\begin{equation}
    \mathcal{T}^{(J)}(J) = \begin{bmatrix}
        \Omega^{-n/2} & 1/2 \\
        -1/2 & J^n \Omega^{-n/2} - \Omega^{n/2}/4
    \end{bmatrix},
\end{equation}
one can observe that
\begin{multline}
    P_{2L-1}(\Omega^{-n}) = \begin{bmatrix}
        \Omega^{-n/2} & 1/2
    \end{bmatrix}
    \mathcal{T}^{(h)}(h_L) 
     \mathcal{T}^{(J)}(J_{L-1}) \dots \\
     \dots
     \mathcal{T}^{(J)}(J_{1})\mathcal{T}^{(h)}(h_1)
      \begin{bmatrix}
          1 \\ -\Omega^{n/2}/2
      \end{bmatrix}.
\end{multline}
The operation of the RG comprises repeated multiplications of these matrices, three at a time. The decomposition above has the nice property that, approximately,
\begin{equation}\label{eq:decim}
    \mathcal{T}^{(h)}(h_1) \mathcal{T}^{(J)}(J)\mathcal{T}^{(h)}(h_2) \propto \mathcal{T}^{(h)}\left( \frac{h_1 h_2}{J} \right),
\end{equation}
with corrections that go to zero as $J$ becomes big and $\Omega$ small. An identical relation holds with $h$'s and $J$'s interchanged. 
This means that if $J$ is a bond with transverse fields $h_{1,2} \ll J$ next to it, we can replace the two spins and the bond with a new spin with transverse field $h_1 h_2/J$, so long as we are interested in small single-particle energies compared to these scales (or equivalently the low-frequency contribution to $E_0$, as in \eqref{eq:E0formula}). This is exactly the result one would derive in a more pedestrian approach via second-order perturbation theory.

This implies immediately that strong-disorder RG at criticality leads to an identical infinite-randomness fixed point to the TFIC with, e.g., the same tunnelling exponent $\psi=1/2$~\cite{fisher1995critical}. (This fact could have been deduced by simply inspecting $\varepsilon_k = z_k^{-1/n}$.) Moreover the RG trajectories are identical decimation-by-decimation for a family of models labelled by $n$. This should be exploited to reproduce the results of this paper, but also to calculate distributions of correlation functions~\cite{fisher1995critical, FisherYoung}, which we do not see a way to attack exactly. Strong-disorder RG often renders infinite-randomness critical points more tractable than their clean counterparts; this is certainly such an example, since little is known about correlation functions in the clean chain. 
Also, the RG rule \eqref{eq:decim} works just as well for complex couplings $\{J_j, h_j\}$, with the correctness of the RG now controlled by broad distributions of their magnitudes.

% Also, it should be made explicit that since the RG approximates the single-particle levels, a lot can probably be said about the middle of the spectrum. 

% In our language the RG is a procedure for approximating the random polynomial whose roots give the single-particle energies by a lower-degree polynomial, precisely in the regime of small single-particle energies. Given the importance of finding roots of high-degree polynomials for problems in physics and elsewhere, it would be interesting to understand what classes of random polynomials possess a similar infinite-randomness structure.

\textit{Disguised free fermions.---} There is another class of one-dimensional Hamiltonians introduced by Fendley~\cite{FendleyFreeFermionsInDisguise} which are free fermionic in the sense of having a spectrum of the form \eqref{eq:spectrum} with $n =  2$, despite not being quadratic in Jordan-Wigner fermions. See also generalizations in Refs.~\cite{Graphs2021ElmanChapmanFlammia, Graphs2023ChapmanElmanMann, FendleyPozsgay,PozsgayCircuits}. These models are solved using polynomials which obey recursion relations similar to \eqref{eq:recursion}. A disordered version of Fendley's model was studied by Alcaraz \textit{et al.}~\cite{Alcaraz2023RandomFFID}, numerically and using strong-disorder RG (formulated as perturbation theory). A number of infinite-randomness critical points were found, all of the Ising type. The matrices which generate these recursions are larger than $2\times 2$, so the theory is naively more complicated than that used here, which relied on the consideration of a single random variable of the form $P_{M}/P_{M-1}$~\cite{McCoyWuThermodynamics, DerridaHilhorst1983, NL}. Nevertheless the random couplings appear in these other polynomials in similar combinations~\cite{CalanLuckKesten}, which ought to make some progress possible. It would also be nice to reverse-engineer free models whose single-particle energies are related to other interesting classes of products of random matrices.

\textit{Discussion.---} Let us make two more general comments on subjects which greatly interest us.

It is remarkable that the critical behaviour of the disordered free parafermionic models is independent of $n$. This strange superuniversal behaviour---identical critical exponents for a number of different theories with clearly different symmetries and therefore naively different ``operator content''---seems to exist for other families of infinite-randomness fixed points in one spatial dimension~\cite{SenthilMajumdar}, as well as in higher dimensions~\cite{KangSuperuniversality}.
(Other universality classes are known, however~\cite{HymanYang, Monthus, Refael, DamleHuse, HoyosSUN, HoyosSU2}.) We hope that access to exact results that demonstrate superuniversality of infinite-randomness fixed points will be useful for progress in understanding this phenomenon.

A particularly transparent solution of the 2d square-lattice Ising model with fully inhomogeneous couplings is via a mapping to a dimer problem which is solvable via Pfaffians~\cite{kasteleyn, McCoyWuBook}. The relation between this solution and free fermions can be made explicit~\cite{Samuel}. What the analogues of the 2d Ising model/dimers/Pfaffians are for parafermions is an open question. We believe our Eq.~\eqref{eq:E0formula} will be of value in answering this. 
The link to the so-called $\tau_2$ model~\cite{Baxter1989JStatPhys,Baxter2014tau2, AuYangPerk2014, AuYangPerkII}, effectively associated with a particular frequency of the quantum chain, is tantalizing but not quite what we want. It is at least worth fully understanding the relations between these various models for $n=2$.

\textit{Acknowledgments.---} 
We are grateful to V. Calvera, P. Fendley, J.A. Hoyos, D.A. Huse, C. Murthy and N. O'Dea for discussions. A.P. and A.C. were supported by Stanford Graduate Fellowships.

\bibliography{references}
\end{document}